\documentclass[fleqn,usenatbib]{mnras}

\usepackage[T1]{fontenc}

\DeclareRobustCommand{\VAN}[3]{#2}
\let\VANthebibliography\thebibliography
\def\thebibliography{\DeclareRobustCommand{\VAN}[3]{##3}\VANthebibliography}

\usepackage{graphicx}	
\usepackage{amsmath}	
\usepackage{amssymb}	
\usepackage{xcolor}	 
\usepackage{subcaption}

\newcommand{\afe}{$[\alpha/\rm{Fe}]$}
\newcommand{\feh}{$[\rm{Fe}/\rm{H}]$}
\newcommand{\msun}{M$_\odot$}
\newcommand{\bpass}{{\sc{bpass}}}

\title[$\alpha$-enhancement in young populations]{The Dependence of Theoretical Synthetic Spectra on $\alpha$-enhancement in Young, Binary Stellar Populations}

\author[Byrne et al.]{
C. M. Byrne,$^{1}$\thanks{E-mail: conor.byrne@warwick.ac.uk (CMB)}
E. R. Stanway,$^{1}$
J. J. Eldridge,$^{2}$
L. McSwiney,$^{1}$
and O. T. Townsend$^{1}$
\\
$^{1}$Department of Physics, University of Warwick, Gibbet Hill Road, Coventry, CV4 7AL, UK\\
$^{2}$Department of Physics, University of Auckland, Private Bag 92019, Auckland, New Zealand
}

\date{Accepted XXX. Received YYY; in original form ZZZ}

\pubyear{2022}

\begin{document}
\label{firstpage}
\pagerange{\pageref{firstpage}-\pageref{lastpage}}
\maketitle

\begin{abstract}
The enhancement of $\alpha$ elements such as oxygen is an important phase in the chemical evolution of the early Universe, with nebular material becoming enriched in these elements sooner than iron. Here we present models which incorporate stellar spectra with $\alpha$-enhanced compositions, focusing on the impact on the integrated light of young stellar populations, including those with large binary star fractions using the Binary Populations and Spectral Synthesis (\bpass) framework, while using Solar-scaled stellar evolution models. We find that broad spectrum outputs such as  production of ionising flux, the ultraviolet spectral slope and optical colours are only weakly affected by a change in \afe. A number of features such as ultraviolet line indices (e.g. at 1719 and 1853\,\AA) and optical line indices (such as MgB) are sensitive to such changes in composition for a continuously star-forming population and a single starburst population respectively. We find that at ages of more than 1\,Gyr, $\alpha$-enhanced stellar populations appear bluer than their Solar-scaled counterparts, and show expected sensitivity of optical line indices to composition, in agreement with previous work. The ultraviolet stellar absorption lines are relatively insensitive to subtleties in the abundances ratios, although with sufficient measurement precision, a combination of UV line indices may enable a simultaneous measurement of total metallicity mass fraction and \afe\ in young stellar populations. The output models are designated as \bpass\ v2.3 and made available to the community with the aim of assisting interpretation of observations of high-redshift galaxies with the James Webb Space Telescope.
\end{abstract}

\begin{keywords}
binaries: general -- stars: abundances -- galaxies: stellar content -- methods: numerical
\end{keywords}



\section{Introduction}

The chemical evolution of the Universe does not happen uniformly across all elements \citep[e.g.][and references therein]{Kobayashi20}. Elements produced through the $\alpha$ process are synthesised in large quantities in massive stars and core collapse supernovae, while iron group elements like Fe and Ni are produced more readily during Type Ia supernovae. These occur at the end of the life of less massive stars, introducing a long delay time from initial star formation. Consequentially, stars in the early Universe form from material which has a higher $\alpha$ element to iron abundance ratio than stars like the Sun which formed later.

Analysis of deep spectroscopy of distant galaxies has confirmed this expectation, with significant levels of $\alpha$-enhancement observed in galaxy populations at $z\sim2$ \citep{2016ApJ...826..159S,2021arXiv211106410S} and $z\sim3.4$ \citep{2021MNRAS.505..903C}. The advent of the James Webb Space Telescope (JWST) extends such analyses to higher redshifts, and such observations make plain the need for stellar population models which incorporate this property. In order to be of practical benefit for understanding the young stellar populations in the galaxies that JWST will observe, these models will also need to include massive stars, which will be a major contributor to the stellar flux, and the effects of binary stellar evolution, as massive stars are frequently found in multiple star systems \citep{2012Sci...337..444S}.

Most stellar population synthesis models, which are presently used to fit and determine the properties of galaxies at high redshifts, rely on models which scale the overall metallicity mass fraction, $Z$, in proportion to an assumed Solar abundance profile. At low metallicities, this assumption is clearly not valid, and measurements of the gas phase metallicity (typically measured by the strength of emission lines of oxygen and other $\alpha$ elements) and the stellar metallicity (typically measured from iron absorption lines) can produce significantly different results in low-metal environments.

The enhancement of $\alpha$ elements has an impact on all aspects of stellar population synthesis. It affects both the overall evolution of the stars as well as their resulting spectra. Potentially it may also affect the occurrence and outcome of binary interactions. Stellar evolution models of stars with $\alpha$-enhanced compositions has been computed by numerous groups \citep[e.g.][]{Vandenberg14,Pietrinferni06,Pietrinferni21}. These studies have tended to explore old stellar populations in nearby globular clusters, since these are the environments in which $\alpha$-enhancement is seen in the local Universe. Thus the focus of the work was on understanding the effects of $\alpha$-enhancement on the evolution of low mass stars, which have long evolutionary timescales. For example, the recent work by \citet{Pietrinferni21} calculated a grid of models with $\alpha$-enhancement using the BaSTI stellar evolution code. This grid contained only stars which evolve in isolation from any companion and has grid steps of 1\,M$_\odot$ in the initial mass range 10-15\,\msun. No higher mass models have been calculated by these authors, or any others we are aware of. This information is of limited benefit for studying stellar populations in the early Universe, where the flux output is dominated by hot, young, massive stars, which can have initial masses extending well above 100\,\msun, and which are overwhelmingly likely to undergo binary interactions during their lifetimes. These early galaxies are certainly metal-poor, and thus highly likely to also be $\alpha$-enhanced. These are effectively the same stellar populations seen in local old systems, but now observed directly in their infancy.

The effects of $\alpha$-enhancement on the atmospheres and spectra of individual stars has  been studied in some depth, with spectral libraries of non-Solar-scaled compositions found in the literature. These spectral libraries span a range of \feh\ and \afe\ values, but individually have drawbacks for application to  young stellar populations. The theoretical models in the sMILES library of \citet[][K21]{Knowles21} only extend to effective temperatures of $10\,000\,$K, which is insufficient for studying massive stars. The \citet[][C14]{Coelho14} spectral libraries provide spectra up to higher effective temperatures ($30\,000\/$K), but the number of composition variations is comparatively small, with 8 different values of \feh\ at a fixed \afe \ of $+0.4$. The high resolution spectral libraries of \citet[][AP18]{AllendePrieto18Paper} has good parameter space coverage  while also providing a wide variety of compositions, but also extends to only $30\,000$\,K. These $\alpha-$enhanced spectra are calculated for $\lambda>2000$\,\AA, and so do not extend into the far-ultraviolet region of the spectrum, which dominates the production of nebular emission in star-forming galaxies and is the most straight-forwardly observable spectral region in highly-redshifted galaxies. All three of these libraries make use of model atmosphere parameterisations generated by a version of the {\sc{atlas}} code, {\sc{atlas9}} \citep[][]{Castelli03} in the case of C14 and AP18 and {\sc{atlas12}} \citep[][]{Kurucz05} in the case of K21. There is also a variation in choice of radiative transfer code used to estimate the resultant atmosphere spectra, with C14 using {\sc{synthe}} \citep[][]{Kurucz81}, while AP18 and K21 used the {\sc{ass}}$\epsilon${\sc{t}} code \citep[][]{Koesterke08,Koesterke09} to compute their spectra from a given set of atmospheric parameters. 

Some population synthesis calculations have been done which self-consistently consider both evolution models and spectra that have been $\alpha$-enhanced. \citet{Vazdekis15} carried out a study of old- and intermediate-aged stellar populations at a low spectral resolution (FWHM $2.51$\,\AA). The alpha-enhanced isochrones they used extended only to 10\,\msun. They found that at ages above about a Gyr, an $\alpha$-enhanced population appears bluer at the blue end of the spectrum and redder at red wavelengths than a Solar-scaled mixture. This work focused on the optical spectrum, and did not extend into the ultraviolet. Earlier work by  \citet{Percival09} investigated $\alpha$-enhanced stellar populations in the context of Galactic globular clusters, and such models were able to reproduce the spread of ages and compositions of a large sample of globular clusters in the Milky Way. The high-resolution spectra used in that work \citep[taken from][]{Munari05} have a short wavelength limit of 2500\,\AA, which still does not reach far enough into the ultraviolet to provide a complete picture of the properties and impact of massive stars.

The need for further work is motivated not only by the importance of extending spectral synthesis models to bluer wavelengths and earlier times, but also by a fundamental property of the young stellar populations themselves. A key feature of massive stars in young stellar populations - the role played by binary interactions - has not been taken into account in any of these earlier studies. The majority of massive stars are found in binary systems, with some 70\,per cent of such stars interacting during their evolutionary lifetime \citep{2012Sci...337..444S}. Stellar evolution models which include binary evolution are rare and computationally expensive but crucial for developing a clear understanding of young stellar populations \citep{Stanway16, 2017PASA...34...58E}. The only stellar population synthesis code to study in detail the evolution of massive stars, including binary interactions, at low metallicity is the Binary Populations and Spectral Synthesis (\bpass) project.

Here we outline preliminary work done to incorporate $\alpha$-enhanced stellar spectra into the spectral synthesis portion of Binary Populations and Spectral Synthesis \citep[\bpass,][]{2017PASA...34...58E}. We highlight some of the key differences that $\alpha$-enhanced spectra make to the spectral synthesis outputs\footnote{We restrict this analysis to a single spectral library. The detailed variation in effects resulting from the choice of spectral library on simple stellar populations in a binary population synthesis will be explored in a future publication, as it is not the main objective of this work.}, identify features which are good indicators of $\alpha$-enhancement, and make the output data available to the community as a limited release version numbered v2.3. 

\section{$\mathbf{\alpha}$-enhanced Spectral Synthesis}

\subsection{Population and Spectral Synthesis}

In the local Universe, detailed observation and analysis of the stellar population is often used to study globular clusters. These are old stellar populations, which no longer undergo star formation. As these are comprised of old, low-mass stars, they form in similar conditions, and can be expected to have similar compositions, to the young stellar populations observed in the early Universe. These old stellar populations in the local Universe can be considered as the low-mass remnants of the young stellar populations in the distant Universe.

As stars in the distant Universe can not be observed individually, our understanding of the stellar content of distant galaxies comes almost entirely from the combined light of the stellar population. Consequentially, the best theoretical tools for comparison with observations come from stellar population synthesis and subsequent spectral synthesis. By generating a stellar population based on a pre-determined initial mass function and an assumption of the star formation history of the galaxy, the evolution of an ensemble of stars can be computed. Then at defined points in time, the relative contributions of each of these stars are combined to produce a composite spectrum of the stellar population. 

Young stellar populations are dominated in flux by many short-lived massive stars, which are well known to have a high multiplicity fraction. Thus it is crucial to consider the effects of binary interactions when making models of young stellar populations. This substantially increases the number of stellar evolution models required to generate a representative stellar population as a distribution of binary star mass ratios and initial separations need to be included in the population synthesis, as well as the considerations of the initial mass function and the star formation history.

\subsection{Choice of metallicity scaling}
\label{sec:zscale}


When dealing with models of non Solar-scaled mixtures, it is important to consider how best to compare these to models with Solar-scaled mixtures. One method is to keep \feh\ fixed while changing \afe\, which would compare models with identical iron content, but a different total $Z$. Another approach is to keep the total $Z$ fixed by choosing an appropriate \feh\ to correspond to the change in \afe. We take the latter of the two approaches in this work. This is also important if one is considering stellar evolution models and stellar spectra with different mixtures.

In the case of stellar evolution models, previous studies such as \citet{1993ApJ...414..580S} have suggested that $\alpha$-enhanced single stellar populations at a given value of \feh\ and \afe\ show isochrones consistent with a Solar-scaled mixture with a overall metallicity, $[M/\rm{H}]$ that satisfies the relation $[\rm{M}/\rm{H}] = [\rm{Fe}/\rm{H}]\log(0.638\,f_\alpha + 0.362)$, where $f_\alpha$ is the linear $\alpha$-enhancement factor.
This scaling relation is consistent with keeping the total metallicity mass fraction constant, given the assumed Solar mixture of the model set. This scaling relation has not yet been tested in binary stellar populations or in the \bpass\ stellar evolution code, and the precise values of the numerical constants need refining to account for changes in our understanding of the Solar composition since the work of \citet{1993ApJ...414..580S}.

Recent modelling of massive stars with a selected non-Solar-scaled composition has been carried out by \citet{Grasha21}. They find that at Solar metallicity, evolutionary tracks of massive stars are not strongly influenced by composition (only total metallicity), although the role of composition becomes more noticeable at low metallicities for stars with a mass greater than 50\,\msun.

\begin{figure}
    \centering
    \includegraphics[width=0.48\textwidth]{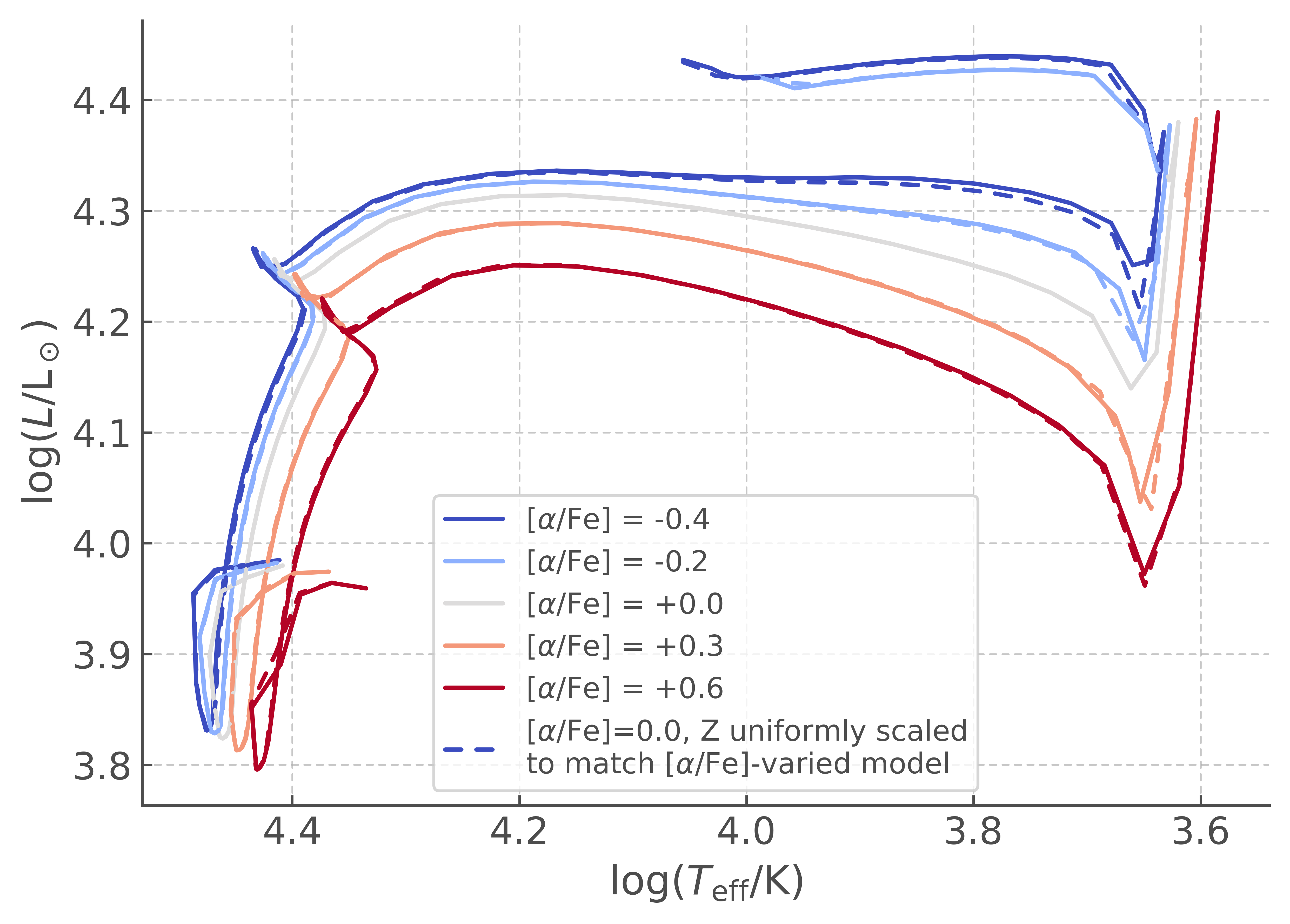}   
    \caption{Hertzsprung-Russell diagram showing the evolution of a 10\msun\ star for a variety of compositions, starting with $Z=0.002$ as a base metallicity. The solid lines represent tracks of models with a fixed iron abundance and $\alpha$-enhancements as indicated by the legend, while the dashed lines show models which have been scaled uniformly from the Solar composition to match the total $Z$ of the corresponding $\alpha$-enhanced model.}
    \label{fig:hr}
\end{figure}

In this preliminary work we proceed based on the assumption that the evolution of a star's mass, luminosity, temperature are predominantly sensitive to the overall metallicity mass fraction rather than the relative abundances of the individual heavy elements. Trial stellar evolution calculations at moderate stellar masses with {\sc{mesa}} \citep[][]{Paxton19} indicate that this assumption is reasonable. Main Sequence lifetimes, stellar radii and peak luminosities are very similar for $\alpha$-enhanced compositions when the overall metallicity fraction is scaled in proportion to the solar abundance to match that of an $\alpha$-enhanced model. Fig~\ref{fig:hr} demonstrates this, showing Hertzsprung-Russell evolution tracks of a 10\,\msun\ star given a fixed iron abundance and 5 different values of \afe, shown by the solid lines. The dashed lines show the evolution of an equivalent star, with the total metallicity scaled uniformly to match the total metallicity of its $\alpha$-enhanced counterpart. The close similarity between the solid and the dashed lines indicates that the overall evolution of the star appears to be most sensitive to the total metallicity, rather than either the Fe or $\alpha$ element abundances considered in isolation. For this reason, we will focus on matching stellar spectra to stellar models based on their total metallicity, regardless of their composition. More detailed study of the effects of $\alpha$-enhancement on stellar evolution models at moderate and high masses is required to ascertain if this is the most reasonable approach to take over the entire range of mass and metallicity. Studies such as \citet{Farrell21} are now starting to explore the effects of $Z$ and CNO abundances on stellar evolution at high masses, along with the impacts of other sources of uncertainty, such as convection physics and stellar rotation.
It is important for later sections of this paper to note that by matching alpha-enhanced spectra to stellar models according to their total $Z$, an $\alpha$-enhanced spectrum will be iron-poor relative to a solar-scaled mixture with the same value of $Z$.

\subsection{\bpass\ stellar evolution models}
\bpass\ is a suite of population and spectral synthesis models which traces the evolution of a simple stellar population from 1 Myr to 100 Gyr in age \citep{2017PASA...34...58E, 2018MNRAS.479...75S}. The synthesis uses a grid of detailed stellar evolution models incorporating both single star and binary evolution pathways, built using a custom version of the Cambridge {\sc{stars}} code. The output of individual stars are combined using a \citet{2001MNRAS.322..231K} initial mass function which extends from 0.1 to 300\,M$_\odot$. They are also weighted by empirical estimates of the mass-dependent binary fraction, period distribution and mass ratio distribution \citep{2017ApJS..230...15M}. The stellar evolution models are matched to appropriate stellar atmosphere models by temperature, surface gravity and surface composition in order to predict the integrated UV-IR spectrum of the stellar population and its photometric colours. As explained above, the inclusion of binary pathways and hot star spectra is particularly important when considering young populations.

BPASS stellar models assume an initial uniform composition of the stars such that the initial hydrogen abundance by mass fraction is given by $X = 0.75 - 2.5Z$. 
Here $Z$ is the initial metallicity mass fraction which scales all elements from their Solar abundances as given in \citet{Grevesse93} in which $Z_\odot=0.020$. The nucleosynthesis and radial distribution of H, He, C, N, O, Ne, Mg, Si and Fe are tracked at each time step in the detailed stellar models. For more details on the input physics and construction of the \bpass\ stellar models, see \citet{2017PASA...34...58E}. 

\subsection{C3K Spectral Libraries}

In this work we extend the BPASS spectral libraries to include Main Sequence and Giant Branch populations from the C3K model grids, computed using the {\sc{atlas12}} and {\sc{synthe}} programs \citep{Kurucz70,Kurucz93,Kurucz81} and the latest Kurucz linelists, matching those incorporated into the Flexible Spectral and Population Synthesis \citep[FSPS,][]{Conroy09,Conroy10} models, as provided to us by C. Conroy and B. Johnson (priv. comm.). The provided spectra have a native resolution of R $=10\,000$ and cover a range of compositions from $-4.0\le[\rm{Fe}/\rm{H}]\le0.5$ (in steps of $0.5$ for $[\rm{Fe}/\rm{H}]\le-3.0$ and $0.25$ for $-3.0\le[\rm{Fe}/\rm{H}]\le0.5$) and $-0.2\le$\afe$\le0.6$ (in increments of $0.2$) These abundance ratios use the Solar mixture of \citet{Asplund09} as their reference values. The spectra cover a wavelength range of $1000\le\lambda/$\,\AA$\le20\,000$, and are supplemented by an approximate flux at $100\le\lambda/$\,\AA$\le1\,000$ calculated via summing the estimated line opacities from {\sc{synthe}} output, smoothed with an R$=250$ kernel. There is a broad coverage of effective temperatures ($2\,500\le T_{\rm{eff}}\le50\,000$), with surface gravities between $-1\le\log(g/\rm{cm}\,\rm{s}^{-2})\le5.5$. at low temperatures. Low-gravity coverage tapers off at higher effective temperatures. More detail on the construction of these models can be found in Sections 3.3 and 5.4 of \citet{Choi16}.

\subsection{Final spectral grids}

To incorporate $\alpha$-enhanced spectra into \bpass, we interpolate between the C3K spectral grids in \feh\ at fixed \afe\ to produce models which match each of the 13 metallicity mass fractions that correspond to the underlying \bpass\ stellar evolution models. When each value of \afe\ is accounted for, this creates 65 grids of spectra. These are then smoothed and resampled at a fixed wavelength separation of 1\,\AA\ for use in the \bpass\ spectral synthesis routines. As discussed above, this metallicity matching approach does produce a discrepancy between the composition of the stellar evolution models and the corresponding spectra. This is an acceptable compromise if one assumes that the stellar evolution is mostly sensitive to the overall metallicity rather than the specific distribution of metals, as suggested by our trial calculations in Section~\ref{sec:zscale}. 
To produce the most self-consistent results, the calculation of stellar evolution models with compositions that match the spectra should be carried out and this will be treated as a priority in the further development of the capabilities of \bpass.

\begin{figure}
    \centering
    \includegraphics[width=0.45\textwidth]{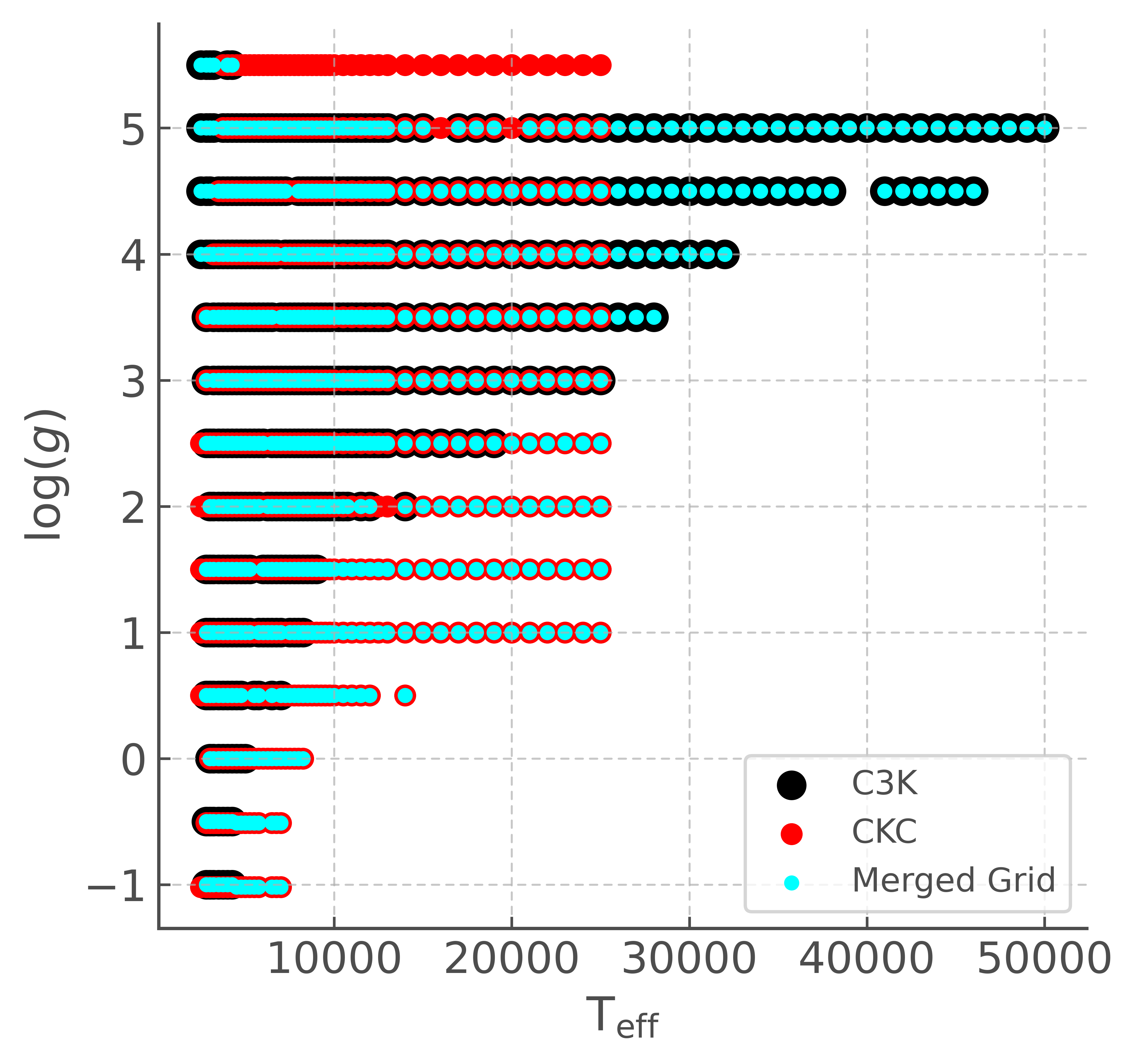}   
    \caption{Temperature and surface gravity coverage (at $Z=0.02$, \afe=0.0) of the new C3K model grid, the CKC grid used in \bpass\ v2.2.1, and the merged grid used in this work (v2.3) which supplements the C3K grid with some CKC models in the high-temperature, low-gravity regime.}
    \label{fig:grid}
\end{figure}

Each interpolated grid comprises approximately 450 stellar spectra for every \{$Z$,\afe\} pair, spanning a temperature range of $2\,500\le T_{\rm{eff}}\le50\,000$ and a surface gravity range of $-1.0\le\log(g/\rm{cm}\,\rm{s}^{-2})\le5.5$, with the lower limit of surface gravity increasing as the effective temperature increases. To maximise parameter space coverage at higher temperatures in the low gravity regime, each grid was supplemented with additional spectra from the existing \bpass\ CKC spectral grid \citep[described in][]{2017PASA...34...58E,2018MNRAS.479...75S}, giving a merged set of models. These supplementary models are not $\alpha$-enhanced, but this region of parameter space is occupied by hot AGB and supergiant stars, which are not expected to be particularly sensitive to $\alpha$-enhancement. Indeed, inspection of the hottest spectra in the existing grid suggest little evidence for strong $\alpha-$dependent absorption lines at high temperatures. Hence this was deemed a more robust approach than extrapolating too far outside of the C3K parameter space. Such an approach supplements each grid with another $\sim$100 models, with the exception of the $Z=10^{-5}$ grids, which do not have a corresponding CKC grid to supplement the C3K models. An example of the temperature and surface gravity coverage of one of these grids ($Z=0.02$, \afe=+0.0) is shown in Figure~\ref{fig:grid}. The spectra used in \bpass\ v2.2.1, labelled as `CKC' are shown as red points, while the new C3K spectra are shown as black points. The combined grid used in this work, indicated by the cyan points, includes spectra from the CKC grid at surface gravities $<3$ which have effective temperatures greater than the hottest models in the C3K grid.

\begin{figure}
    \centering
    \includegraphics[width=0.4\textwidth]{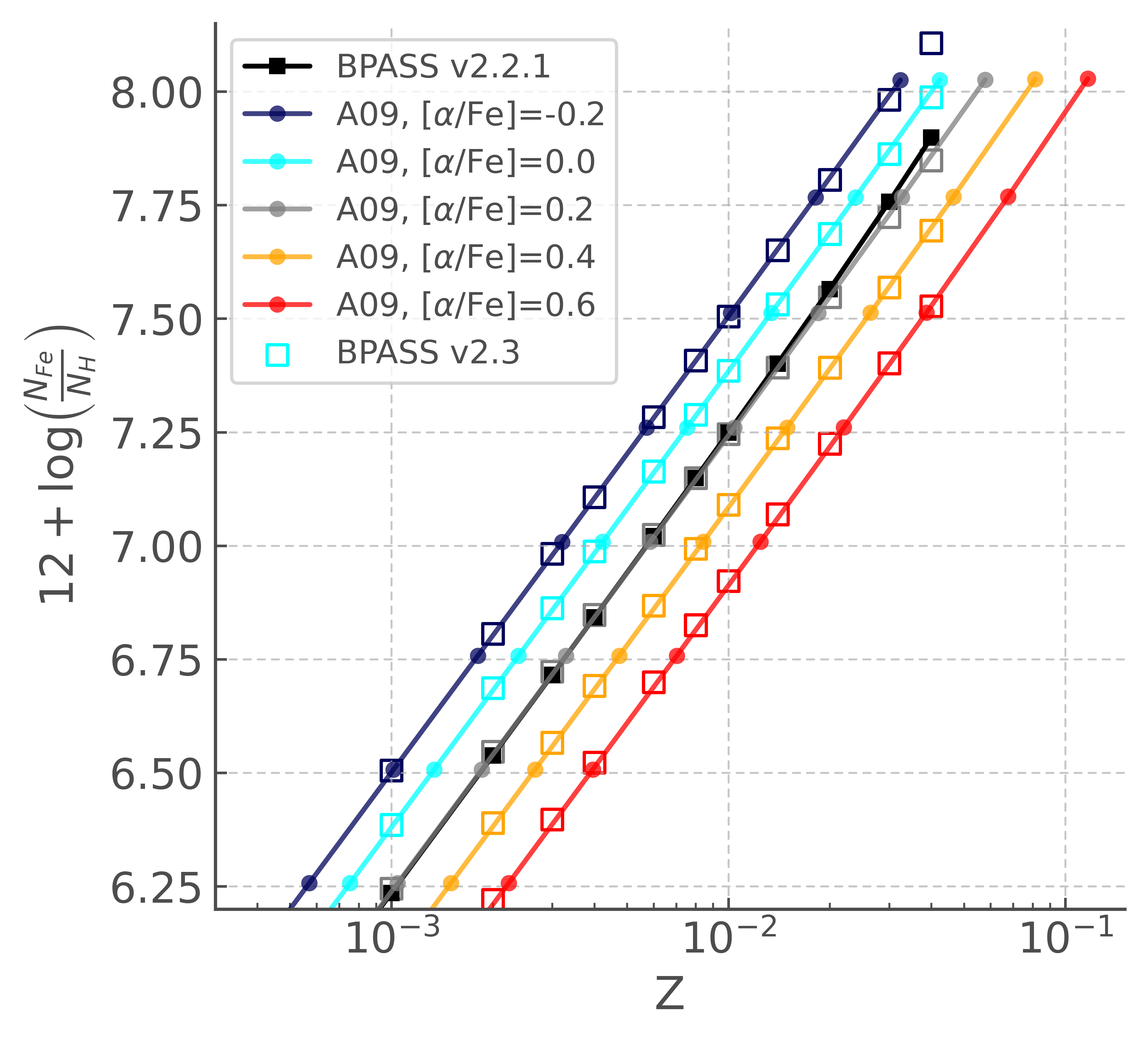}
    \includegraphics[width=0.4\textwidth]{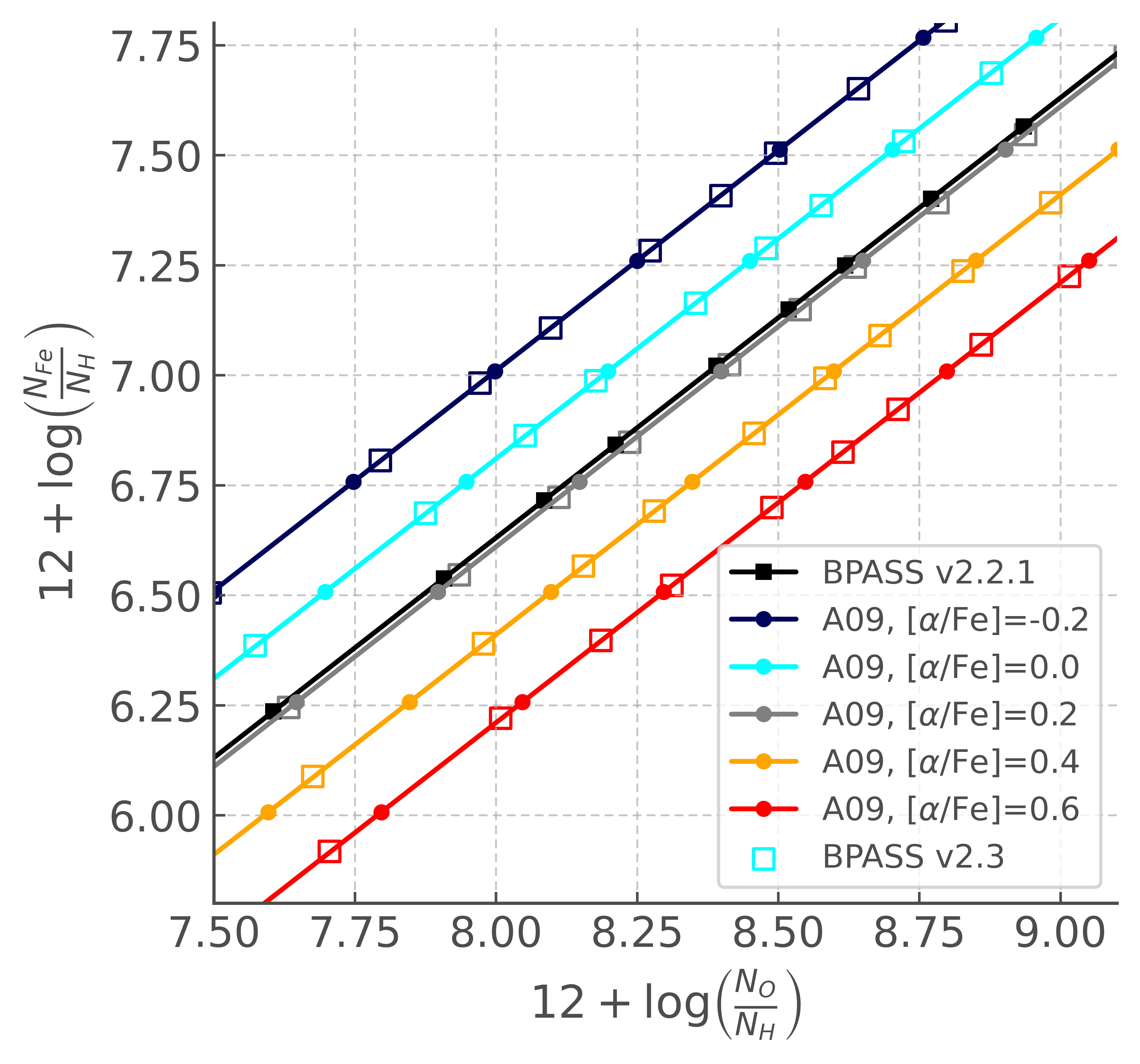}
    \caption{Upper panel: Relationship between the iron-to-hydrogen number ratio, \feh, and total metallicity mass fraction, $Z$ for each of the grids of stellar spectra used in \bpass\ v2.2.1 and this work. A09 refers to the composition profile of \protect\citet{Asplund09}, used in the C3K spectra, with the round points indicating the composition of each spectral grid, while the open squares indicate the position of the interpolated grids used in BPASS v2.3, which match the total metallicity of the underlying \bpass\ stellar evolution models. Lower panel: Relationship between the oxygen and iron abundance ratios for each grid of spectra, with the lines, colours and symbols having the same meaning as the previous panel.}
    \label{fig:feh_Z}
    \label{fig:oh_Z}
\end{figure}

Figure~\ref{fig:feh_Z} illustrates the relationship between \feh\ and $Z$, and \feh\ and [O/H] for both the most recent full release of \bpass\ (v2.2.1), and the new $\alpha$-enhanced atmospheres added in this work (v2.3), focusing on the more densely populated region of parameter space. The axes in the oxygen-to-iron abundance ratio plot were chosen to match those of  Figure 44 of \citet{2017PASA...34...58E}. The initial C3K spectral grids are indicated by filled circles on each of the lines, the interpolated grids created for this work are shown by the open squares, while the filled black squares indicate the compositions of the BPASS v2.2.1 stellar evolution models. This illustrates the manner in which the spectra have been interpolated, with $\alpha$-enhanced models having a lower value of \feh\ at a fixed $Z$. This also highlights the difference in iron abundance between the underlying \bpass\ evolution models and the new spectra, as at a fixed $Z$, the \bpass\ stellar models have a slightly lower value of \feh\ than the corresponding C3K atmosphere at \afe=0. In both panels we can see that the existing \bpass\ v.2.2.1 stellar models quite closely align with the new \afe$=+0.2$ spectra. This is consistent with the \citet{Grevesse93} abundance profile used in \bpass\ stellar evolution models being more oxygen rich than the more recent \citet{Asplund09} abundances used in the C3K spectra. More importantly, this figure highlights the improved coverage of oxygen-rich, iron-poor environments, which previous \bpass\ outputs fail to reach, and is more representative of abundance patterns seen in the early Universe.

We stress that the underlying stellar isochrones remain fixed in both \feh\ and total metallicity mass fraction while the atmospheres have been varied. The robustness of this approach remains to be tested in future work.

{The spectral data grids used by \bpass\ for the hottest binary products, white dwarfs and Wolf-Rayet or stripped helium stars also remain unmodified. In each case, initial $\alpha$-enhancement is expected to have little or no effect on the stellar spectrum.} The TP-AGB phase within \bpass\ is treated in a simplified model as detailed in \citet{2018MNRAS.479...75S}. The importance of such stars to the output spectrum is reduced somewhat within the interacting binary populations because many of the intermediate mass stars experience binary interactions before second dredge up and thus never experience the expected thermal pulses. However, for those models that are single or in wide enough binaries to experience thermal pulses, the following methods are used. After second dredge-up, (i.e. when the hydrogen and helium burning shells move into close proximity) it is assumed that there is strong third dredge-up during the thermal pulses such that the core mass does not grow. The mass-loss rates of \citet{Schroder05,Schroder07} are then applied and the stellar envelope is removed until the star becomes a white dwarf. This simple arrangement does reproduce most of the features of AGB evolution relevant for our spectral synthesis. At most it slightly over estimates the temperature of the AGB stars during thermal pulses and may slightly underestimate the maximum luminosity and final white dwarf masses.

\subsubsection{Individual spectral lines}
\label{sec:lines}

\begin{figure*}
    \centering
    \includegraphics[width=0.48\textwidth]{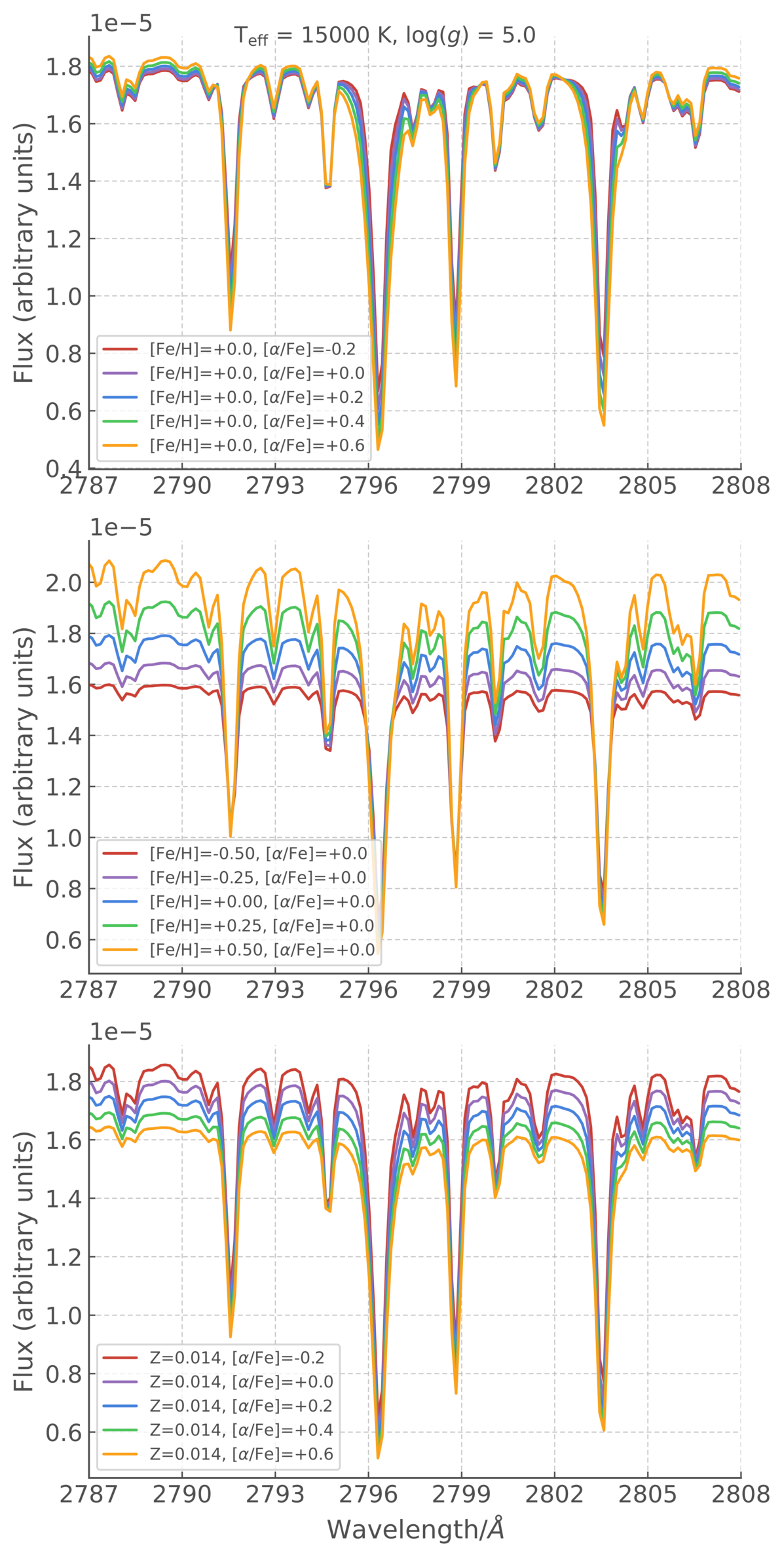}
    \includegraphics[width=0.48\textwidth]{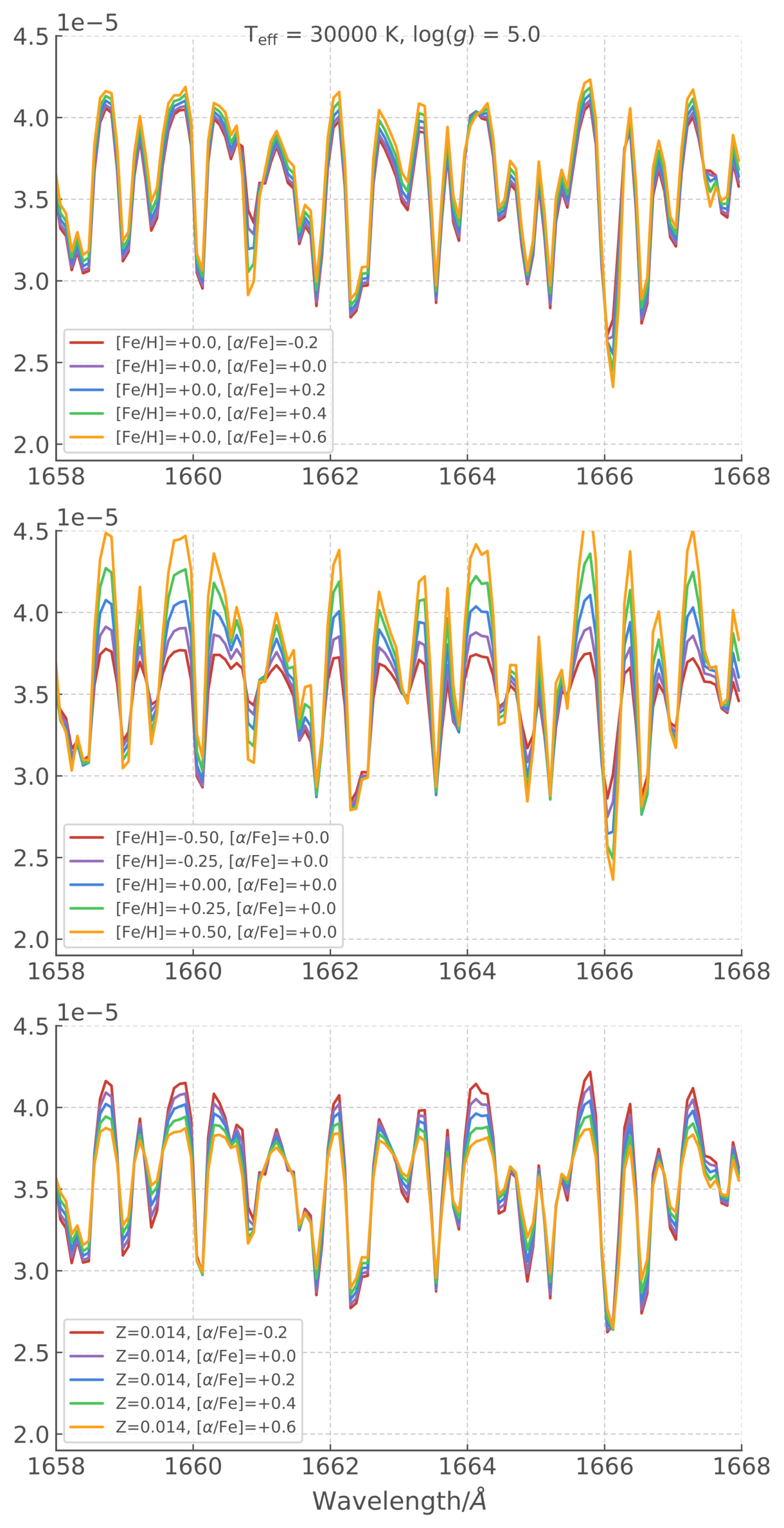}
    \caption{Left: $R=10\,000$ spectra of a $15\,000\,$K star with a surface gravity $\log(g/\rm{cm}\,\rm{s}^{-2})=5.0$ at a variety of compositions, centred on the Mg {\sc{ii}} feature at $\sim$2800\,\AA. Top panel: C3K spectra at fixed \feh=0.0 and variable \afe. Middle panel: C3K spectra at fixed \afe=0.0 and variable \feh. Bottom panel: Interpolated spectra used in this work, with fixed total metallicity $Z=0.014$ and variable \afe. Right: As above, but for a star with effective temperature of $30\,000\,$K and showing the O {\sc{iii]}} doublet at 1660.8, 1666.2\,\AA.}
    \label{fig:3panel}
\end{figure*}

Individual spectral lines are a common method of determining elemental abundances in stars. In the ultraviolet in particular, the spectrum is heavily sensitive to iron line blanketing. Therefore it is instructive to examine some representative stellar spectra and see what effect changes in \feh, \afe\ and $Z$ have on the ultraviolet portion of the spectrum. In Fig~\ref{fig:3panel} we show (at full $R=10\,000$ resolution) spectra for two stars at various compositions, one with an effective temperature of 15\,000\,K and another at 30\,000\,K, both at a surface gravity, $\log(g)=5$. Each set of three panels have matching y-axes for ease of comparison. The top panels show C3K spectral models at a fixed value of \feh $=+0.0$, for each value of \afe. In the central panels, a selection of C3K fixed-$\alpha$ spectra are plotted, ranging over  $-0.5\le$\feh$\le+0.5$. In the lower panel we show the spectra which have been interpolated to match a total $Z$ of 0.014, the closest BPASS metallicity mass fraction to the Solar metallicity of the C3K spectra. The spectrum of the 15\,000\,K star focuses on the Mg{\sc{ii}} doublet at $\sim$2800\,\AA, while the spectrum of the 30\,000\,K shows the O{\sc{iii]}} doublet at $\sim$1663\,\AA\ in the far ultraviolet. In both cases, the effects of changing \afe\ at fixed \feh\ is apparent, with the line depths increasing with increasing \afe. In the case of the Mg{\sc{ii}} lines, it can also be seen that the wings get broader. At fixed \afe, the effects of changing \feh\ can be seen to be much more pronounced, with a significant shift in the location of the continuum, but with line depths which are broadly comparable. Thus, there are two competing effects, with \afe\ serving to increase the line depths, and \feh\ modifying the height of the continuum. As a result, when we compare models at a fixed $Z$ and varying \afe\ (and thus also changing \feh) in the bottom panels, it can be seen how these two effects work in opposite directions, leading to line depths which are very similar, but a continuum which varies in strength, with $\alpha$-enhanced (and iron-poor) spectra having a stronger continuum flux.

\section{Model Spectral Outputs}

\subsection{Global properties}
\subsubsection{UV flux output}
\begin{figure}
    \centering
    \includegraphics[width=0.48\textwidth]{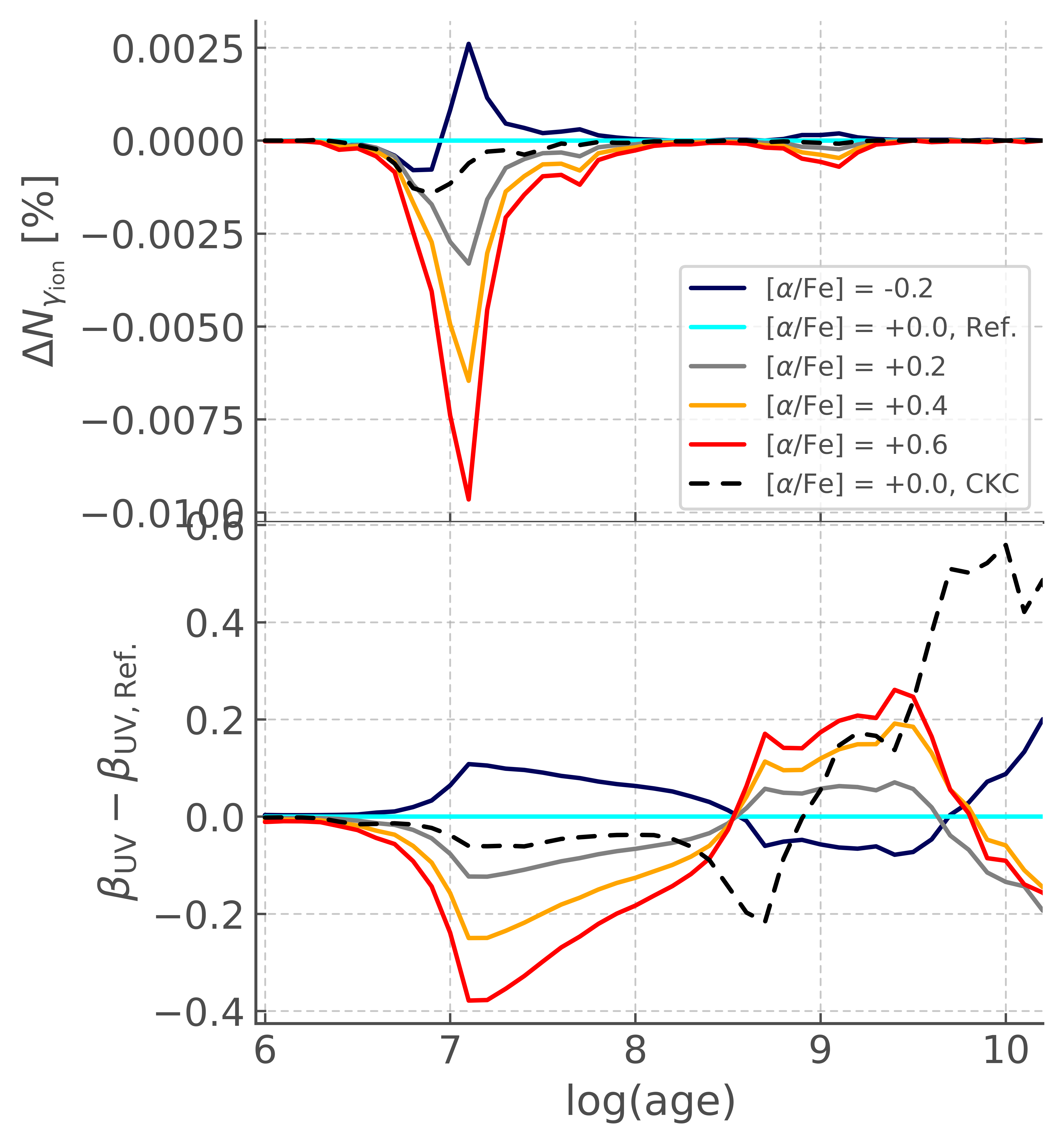}   
    \caption{Upper panel: Percentage change in the rate of production of ionising flux due to $\alpha$-enhancement as a function of stellar age, for $Z=0.002$, with BPASS v.2.3 models at \afe=0.0 being taken as the reference value. The existing \bpass\ v2.2.1 output is included for comparison. Lower panel: Changes to the ultraviolet spectral slope, $\beta$, as a function of age and $\alpha$-enhancement.}
    \label{fig:uvslope}
\end{figure}

To analyse the effects of $\alpha$-enhancement on the composite stellar spectrum, we examine a number of the outputs from the \bpass\ simulations which include the updated spectra. In Figure~\ref{fig:uvslope} we show (for each value of \afe, at a fixed metallicity of $Z=0.002$) the percentage difference in the rate of production of ionising photons (upper panel) and the the difference in the power law index of the UV spectral slope (lower panel) as a function of age, relative to the \afe$=+0.0$ case. This is considered in the case of a single burst of star formation. The previous \bpass\ v2.2.1 models are also included for comparison. We find that the rate of production of ionising photons changes only minimally with different \afe\ values, which is to be expected as the rate is mostly sensitive to the temperature of the underlying stellar models rather than the $\alpha$ elements. The UV spectral slope varies slightly at young ages, with the slope being slightly steeper with increased \afe. Although the spectral slope becomes considerably more $\alpha$-sensitive at late ages ($>1$\,Gyr), this discrepancy appears more significant than it is, given that the overall UV flux is considerably lower at these ages, and the large difference corresponds to a relatively minor change in flux. 

\subsubsection{Colours}

\begin{figure}
    \centering
    \includegraphics[width=0.48\textwidth]{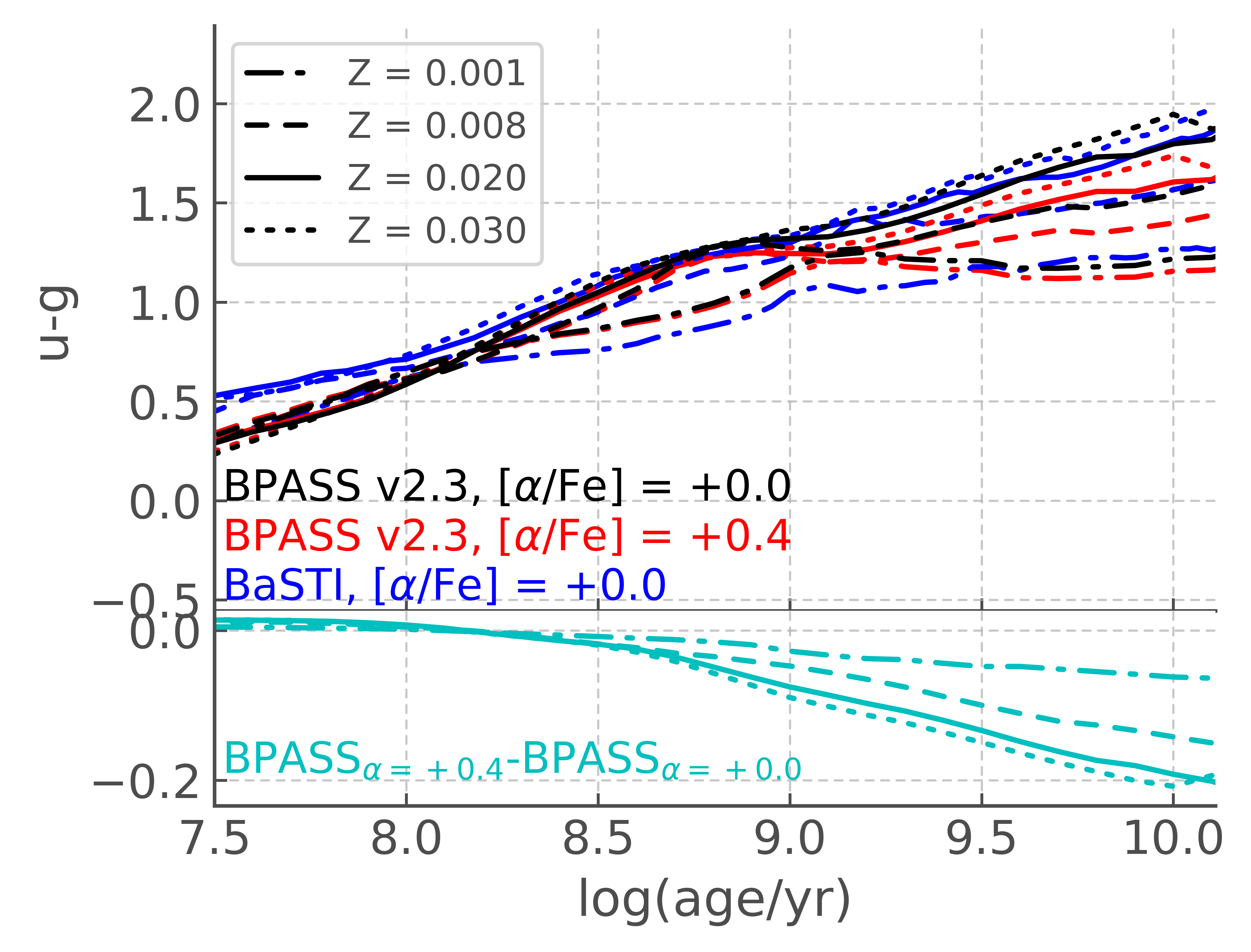}
    \caption{Top panel: Comparison of SDSS $u$-$g$ colours for selected values of metallicity mass fractions between the BaSTI/MILES population synthesis models at `base' metallicity and the \bpass\ v2.3 models with \afe=0 and \afe=+0.4. Lower panel: Colour differences between the \bpass\ v2.3 models with \afe=0 and \afe=+0.4.}
    \label{fig:vazd_comparison}
\end{figure}

Broadband colours are only weakly dependent on the details of the underlying spectra, but nonetheless reflect changes in spectral shape which may be indicative of $\alpha$-enhancement. We compare the predictions of \bpass\ v2.3 to the spectral synthesis results of the Solar-scaled BaSTI population synthesis models combined with MILES spectral models \citep[][]{Vazdekis10,Vazdekis15} in the  $u-g$ colour defined by the Sloan Digital Sky Survey (SDSS) filter profiles. This comparison is shown in Fig~\ref{fig:vazd_comparison} for four different values of $Z$. The time evolution of the $u-g$ colour is broadly consistent between BPASS and the BaSTI/MILES models, covering the same range of values over the same stellar age range. Some minor offsets ($\lesssim0.1$\,mag) can be seen at certain ages, a consequence of the impacts of binary evolution and the different spectral libraries used. In the lower panel we compare the time evolution of $u-g$ colour in the \bpass\ v2.3 models at \afe$=+0.0$ and \afe$=+0.4$ at four different metallicities. At later ages, $\alpha$-enhanced populations appear bluer than their Solar-scaled counterparts, with this effect more pronounced at higher values of $Z$. This behaviour is consistent with that of the $\alpha$-enhanced BaSTI/MILES models \citep[][see the upper right panel of their Figure 19]{Vazdekis15}.

Broadband colours in the Johnson-Cousins UBVRI filters change only slightly with different $\alpha$ enhancements, with a maximum difference of less than 0.05\,mag at a stellar metallicity of $Z=0.02$ in the B, V, R and I bands at all ages, and in the U band for ages less than 1\,Gyr. The larger differences at late ages in the U band indicate that old stellar populations appear bluer when $\alpha$-enhanced at a fixed metallicity, and is in agreement with the SDSS $u-g$ colour results discussed above. 
Comparisons between $\alpha$-enhanced (\afe$=+0.4$) and Solar-scaled mixtures at an age of $\sim12$\,Gyr ($10^{10.1}$\,yr) and a metallicity of $Z=0.004$ show a difference of $\sim0.05$\,mag in the B-V colour, while the V-R colour is not noticeably sensitive to $\alpha$-enhancement. This offset agrees well with that found in previous studies using $\alpha$-enhanced population and spectral synthesis models of single star populations \citep[][their Figure 5]{Cassisi04}.

\subsection{Spectral features}
\subsubsection{Optical line indices}
\begin{figure}
    \centering
    \includegraphics[width=0.48\textwidth]{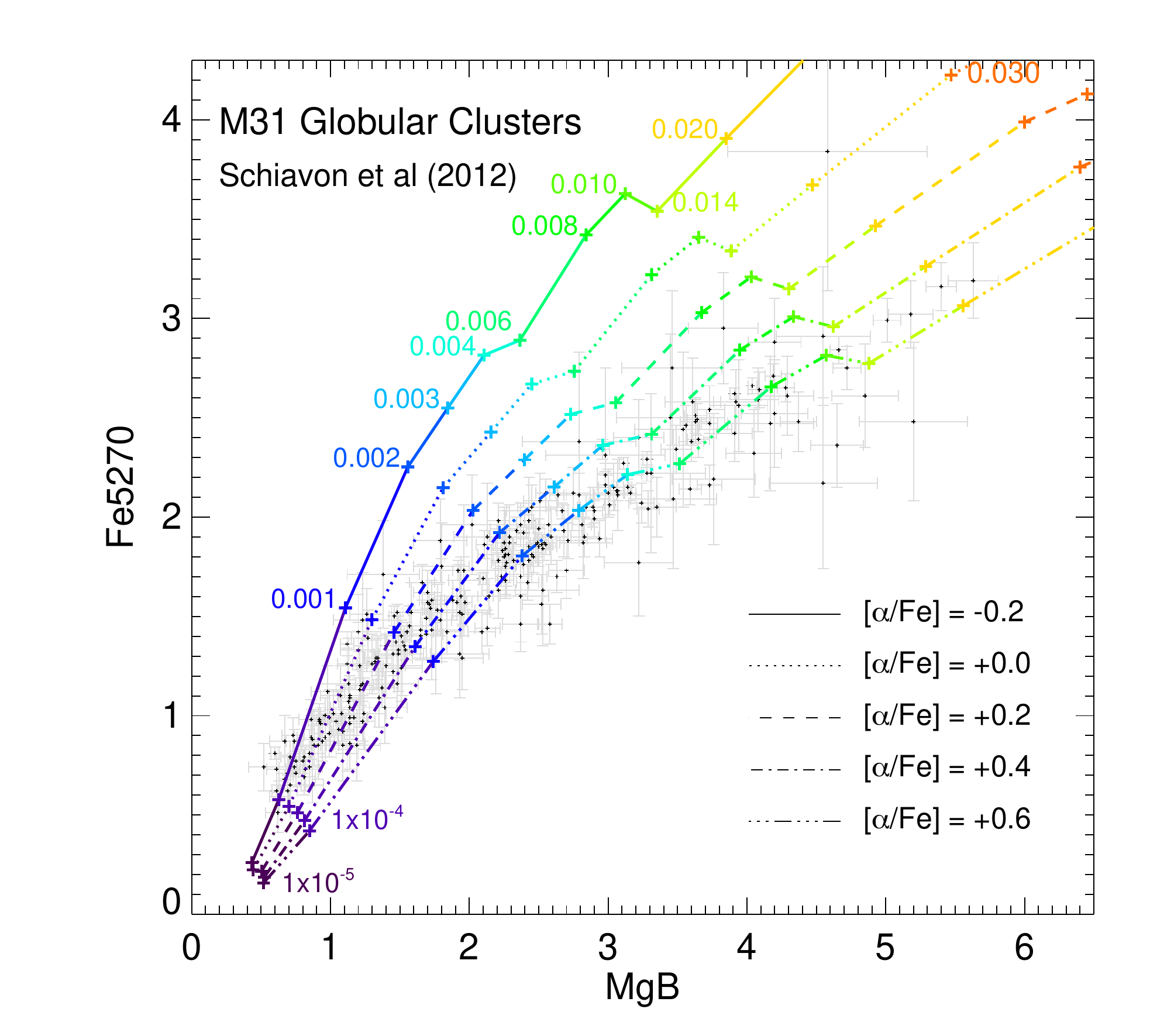}
    \caption{The effect of using $\alpha$-enhanced atmospheres on the optical Lick indices. Here we show the predicted line indices for old stellar populations at log(age/years)=10.0 (10 Gyrs). These two indices are sensitive to total metallicity and $\alpha$-enhancement but weakly age dependent. Metallicity mass fractions are labelled, while different $\alpha$-enhancements are shown with different line styles. For comparison, we also show the line indices measured in the M31 Globular Cluster system by \citet{2012AJ....143...14S}. }
    \label{fig:Lick_Opt}
\end{figure}

While the shape of the spectrum is determined primarily by the underlying stellar models, individual spectral features are known to trace relative elemental abundances. In Figure~\ref{fig:Lick_Opt}, we demonstrate the ability of our new models to recover the expected signature of $\alpha$-enhancement  in stellar populations with ages $>1$\,Gyr, through the optical Lick indices \citep{1997ApJS..111..377W} measured in a sample of M31 globular clusters \citep{2012AJ....143...14S}. These indicate that most of this system has $\alpha$-enhancement factors of between 2.5 and 4 as expected. 

\begin{figure}
    \centering
    \includegraphics[width=0.48\textwidth]{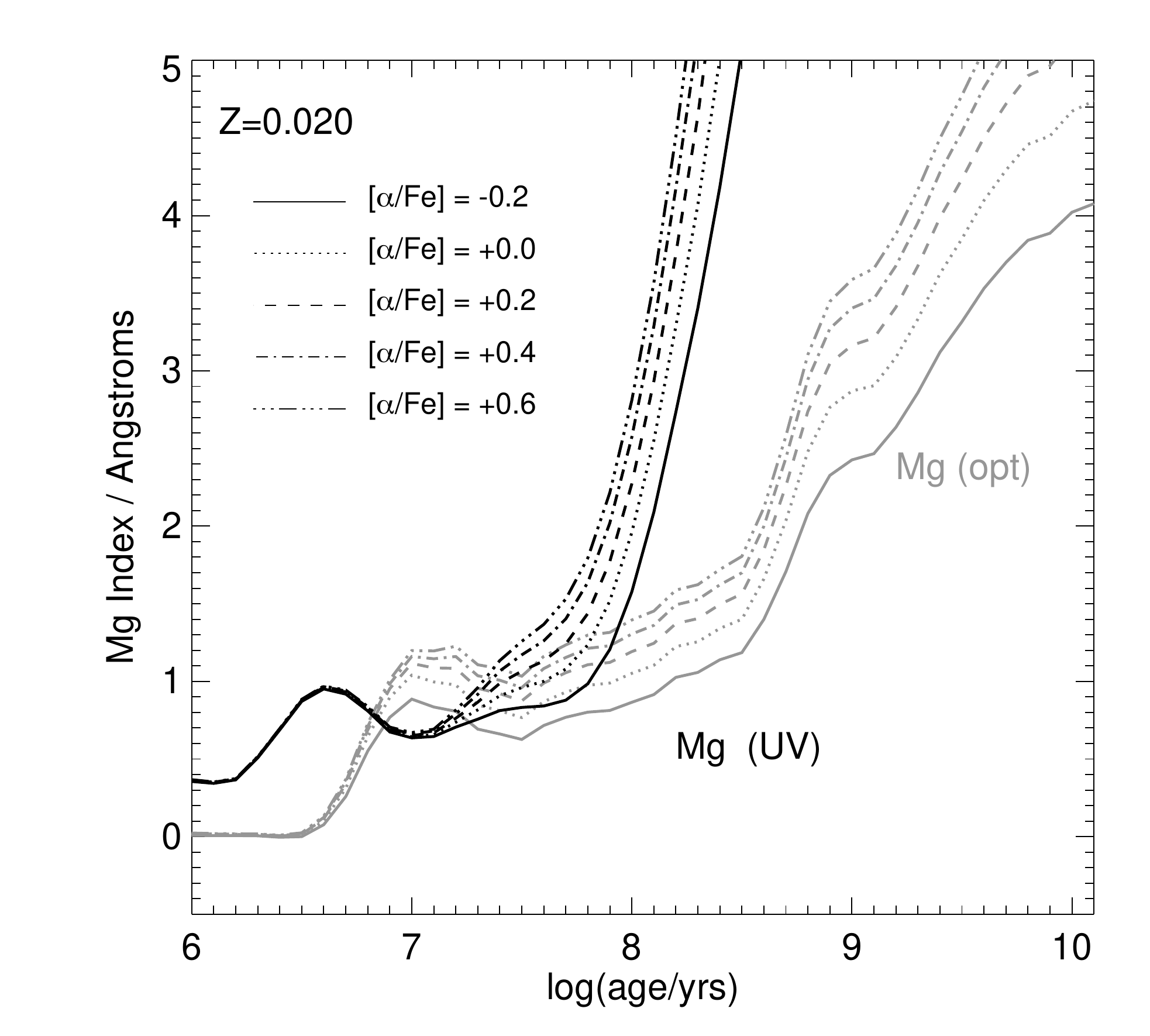}
    \caption{Comparison of the effect of using $\alpha$-enhanced atmospheres on the MgB optical Lick index and the ultraviolet index of the Mg{\sc{ii}} doublet at $\sim$2800\,\AA\ (as defined in this work) as a function of age at $Z=0.020$. The optical line indices are shown by the grey lines, while the UV indices are shown by the black lines. Different $\alpha$-enhancements are shown with different line styles.}
    \label{fig:Opt_vs_UV}
\end{figure}

The time evolution of the optical Mg I absorption complex at $\sim$5180\AA\ is also shown in Fig~\ref{fig:Opt_vs_UV}. This indicates that the absorption grows steadily deeper with increasing age, at fixed $Z$ and \afe, leading to some potential degeneracy in these parameters when the line index is considered alone. We note that the optical line index is of little use in stellar populations with ages less than about a Gyr, since the difference in line strengths with $\alpha$-enhancement is a small fraction of the line depth. This is characteristic of optical line indices, which were developed for use in galaxies in the local Universe.

\subsubsection{Ultraviolet line indices}
\label{sec:uv_index}

In the context of the early Universe, rest-frame ultraviolet measurements will be important, owing to the contribution of short-lived massive stars to the light of star forming galaxies. Fig~\ref{fig:Opt_vs_UV} provides a comparison between the MgB (Mg{\sc{i}}) optical Lick index and the equivalent width of the ultraviolet Mg{\sc{ii}} doublet at $\sim$2800\,\AA, using an index definition of $2785\le\lambda_{\rm{line}}\le2865$, $2725\le\lambda_{\rm{blue\,\,cont.}}\le2745$ and $2870\le\lambda_{\rm{red\,\,cont.}}\le2890$. 
This comparison is shown for a metallicity of $Z=0.020$. While the MgB index is noticeably sensitive to $\alpha$-enhancement across all ages, the Mg\,2800\,\AA\ index is found to be $\alpha$-insensitive at young ages, before the line strength increases significantly at around 100\,Myr. At this point, the overall ultraviolet flux has decreased significantly, and thus it is unhelpful as a diagnostic as a result of large measurement uncertainties.

\begin{figure*}
    \centering
    \includegraphics[width=0.94\textwidth]{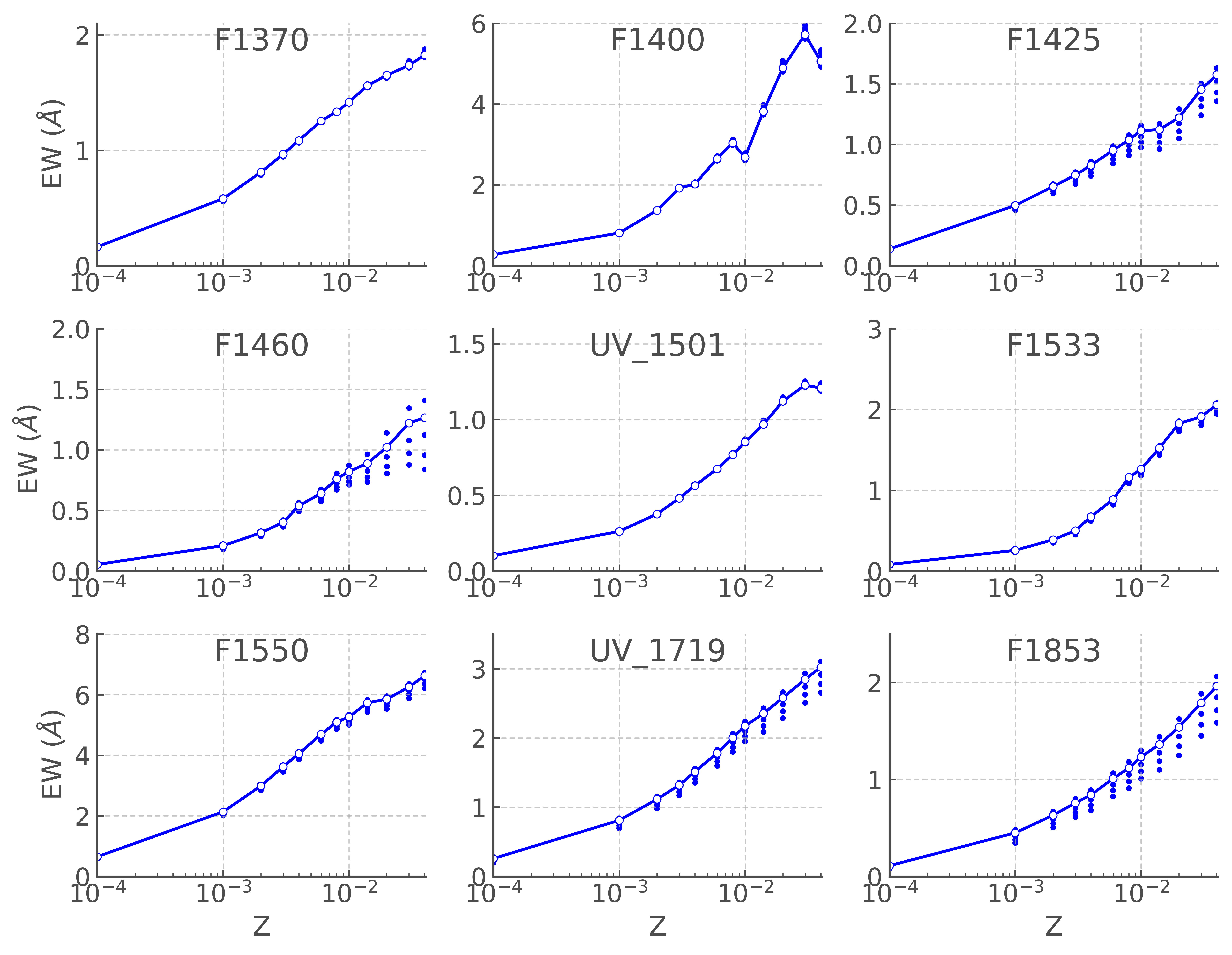}
    \caption{Equivalent width of UV spectral indices from \protect\citet{Calabro21} as a function of metallicity mass fraction, $Z$ for a continuously star forming population of 1\,M$_\odot$\,yr$^{-1}$ for 1\,Gyr. The blue line shows the equivalent width of the indices for the \afe=0.0 case, with the filled blue circles indicating the $\alpha$-enhanced (three points) and $\alpha$-depleted (one point) compositions.}
    \label{fig:calabro}
\end{figure*}

To explore whether there are other features in the UV spectrum which may be more sensitive than the Mg{\sc{ii}} lines, we also measure the strength of a number of other line indices in the UV portion of the spectrum. In this case, we consider a population which is continuously star forming, where there will be a persistent contribution to the UV flux, rather than a single starburst at a fixed age. \citet{Calabro21} have recently employed a set of ultraviolet spectral indices to characterise the metallicity and properties of young stellar populations in the distant Universe. In particular they identified spectral indices at 1501\,\AA\ and 1719\,\AA, as diagnostics which are least affected by external factors such as contamination from the interstellar medium. We evaluate the impact of $\alpha$-enhancement on these indices in Figure~\ref{fig:calabro}, which shows the indices as measured for a stellar population undergoing continuous star formation at a constant rate of 1\,M$_\odot$\,yr$^{-1}$ for 1\,Gyr. We include 9 of the 10 indices discussed in \citet{Calabro21}. The F1978 index is omitted from our analysis as it is found to be largely independent of both metallicity and \afe\ in our models. In these calculations, we have used the level of smoothing and pseudo-continuum window sizes that match those of \cite{Calabro21}.

The blue line with open circles shows the \afe=0.0 models, while filled blue circles indicate different \afe\ values. Note that the y-axis scale for each panel is different. A number of these indices, namely those at 1370, 1400 and 1501\,\AA, show negligible sensitivity to the value of \afe. Additionally, the 1533 and 1550\,\AA\ indices shows only a slight sensitivity. However the remaining four indices (1425, 1460, 1719 and 1853\,\AA) show a more significant sensitivity to composition. These indices generally tend to probe iron features, rather than $\alpha$ elements, which may seem counter-intuitive in the first instance, until one considers that an enhancement of \afe\ leads to a decrease in \feh\ for a fixed value of $Z$. In each of these cases, the equivalent width of these indices is seen to decrease with increasing \afe, indicated by the fact that there are multiple points below the line, and only one (the $\alpha$-depleted composition) above it. This indicates that the change to the continuum produced by a reduction on \feh\ is having a more significant effect than the change in line depths as a result of an increase in \afe. This result is reasonable when one considers that the UV flux of a stellar population which is continuously star forming will be dominated by the youngest, hottest, most massive stars, which are heavily ionised, and very sensitive to the iron line blanketing which affects the continuum. This is also consistent with the behaviour seen in Section~\ref{sec:lines} where the individual stellar spectra showed that at fixed $Z$ hot stars are more sensitive to iron depletion than to $\alpha$-enhancement.

\subsection{Observational Signatures}

\begin{figure*}
    \centering
    \includegraphics[width=0.48\textwidth]{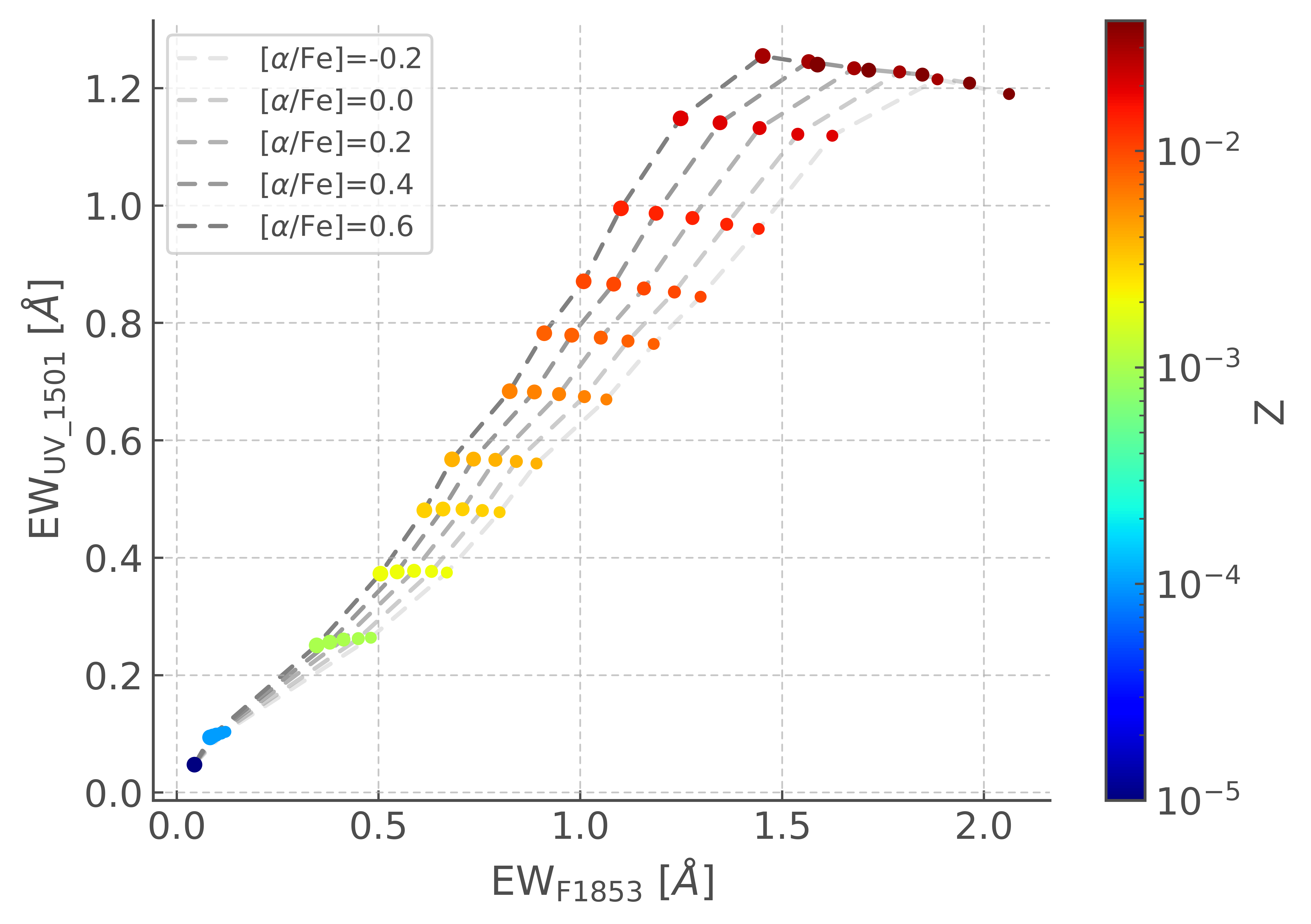}
    \includegraphics[width=0.48\textwidth]{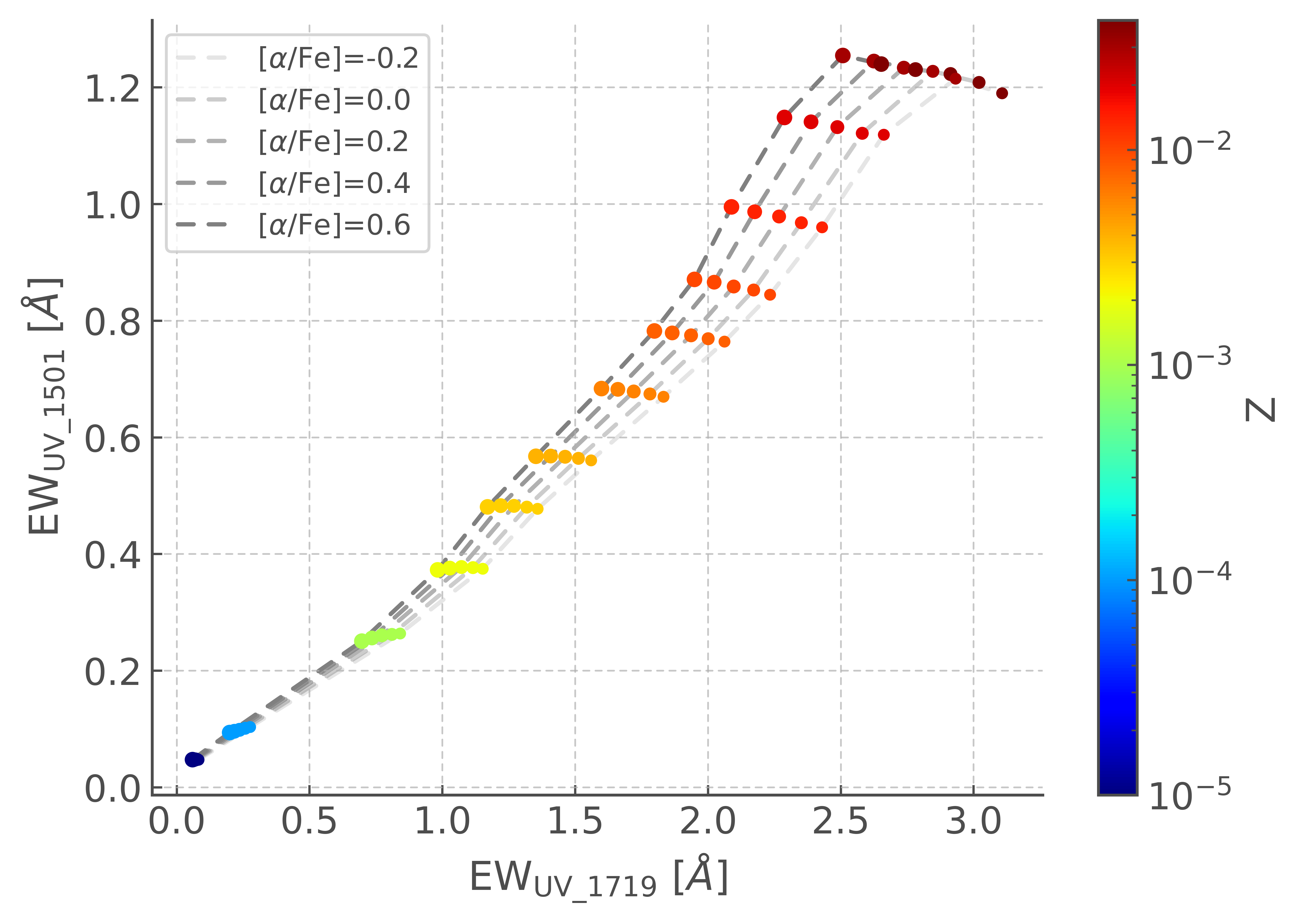}
    \caption{Equivalent width of the 1853\,\AA\, 1719\,\AA\ and 1501\,\AA\ spectral indices, colour coded by $Z$, with the size indicating the magnitude of $\alpha$-enhancement, with larger symbols indicating a higher value of \afe.}
    \label{fig:indexindex}
\end{figure*}

As some indices are $\alpha$- (or iron-) sensitive, while others are only sensitive to total metallicity, measurement of multiple indices could be used to determine $Z$ and \afe\ simultaneously. Fig~\ref{fig:indexindex} shows two such diagnostic diagrams, firstly the 1501 and 1853\,\AA\ indices and secondly the 1501 and 1719\,\AA\ indices. In both cases, measurements of the line indices with a precision of $\sim$0.1\,\AA\ would provide sufficient discriminating power to determine $Z$ and \afe\ with reasonable accuracy. While the $\alpha$-sensitivity of the 1853\,\AA\ index is higher than the 1719\,\AA\ index, the equivalent width is typically smaller, and it is more prone to contamination by the interstellar medium. Consequently, measuring the 1501 and 1719\,\AA\ indices may be the most promising method for determining these properties in stellar populations with continuous star formation.

We note that these models, like those used by \citet{Calabro21}, have been calculated without including nebular emission. Doing so is beyond the scope of this work and requires the use of photoionization radiative transfer codes.
The ionizing spectrum is largely insensitive to $\alpha$-element enhancement (see Fig \ref{fig:uvslope}), and the stellar absorption features are weak in young stellar populations (e.g. age $<1$\,Gyr, see Fig \ref{fig:indexindex}). Thus we expect little $\alpha$-dependence in the nebular emission line or continuum spectrum except where the wavelength coincides with a strong absorption or emission line associated with an $\alpha$-element species \citep[see e.g.][]{2016ApJ...826..159S}.

\section{Future Developments}

It is important to emphasise that only the stellar spectra have been changed in this work. The underlying stellar evolution models still use a Solar-scaled mixture, with the $\alpha$-enhanced spectra being chosen so that they match the total metallicity mass fraction of the underlying evolution models, in accordance with the result of trial stellar evolution calculations with {\sc{mesa}}. The detailed stellar models on which BPASS is built require substantially longer to run than the analytic models used in "rapid" population synthesis models, particularly when the effects of binary interactions are incorporated. 

From the spectral synthesis perspective, a drawback of the current \bpass\ output spectra is their fixed 1\,\AA\ resolution. While this is a reasonable resolution in the rest-frame optical and infrared, it is becoming inadequate in the rest-frame ultraviolet, particularly given the imminent arrival of observations from JWST. Upgrading the spectral resolution of \bpass, together with computing a full grid of non-Solar scaled stellar evolution models for a more complete and self-consistent population synthesis, remains as future work which the authors will undertake in due course.

While the composite spectrum of the stellar population shows relatively minor sensitivity to \afe\ in the UV, optical line indices appear to be considerably more sensitive to this change of composition, as shown in Fig~\ref{fig:Lick_Opt} and Fig~\ref{fig:Opt_vs_UV}. Further analysis of these $\alpha$-enhanced models, in particular looking at the optical spectrum, will form the basis of a future publication. Additionally, by focusing on the optical spectrum, where there are multiple $\alpha$-enhanced stellar spectral libraries (e.g. C14, AP18, K21) publicly available, it will be possible to study and evaluate the uncertainties that the choice of stellar spectra introduce to the process of population and spectral synthesis.

\section{Conclusions} 

In this work we have presented the initial work carried out to incorporate $\alpha$-enhanced stellar spectra into the \bpass\ stellar population and spectral synthesis project, motivated by the need for spectral synthesis models of young stellar populations with a composition that is representative of what is seen in observations in the distant Universe. This was achieved through the integration of Conroy C3K spectral libraries with values of \afe\ between -0.2 and +0.6.
The work presented here represents the first efforts to include $\alpha$-enhanced stellar spectra with broad wavelength coverage in a detailed binary population and spectral synthesis calculation. These models should prove useful with the imminent arrival of observations of the distant Universe by JWST, where stellar populations are young (thus having a high binary fraction) and are expected to have a composition that is enhanced in $\alpha$ elements.

In terms of ultraviolet flux, these models indicate that the rate of production of ionising photons is largely insensitive to \afe, with insignificant variation seen. Broadband colours behave in a manner consistent with that seen in the literature, with older stellar populations appearing bluer when $\alpha$-enhanced, while the colour of young stellar populations is effectively unchanged. At a metallicity of $Z=0.001$, the difference in  $u-g$ colour between a Solar-scaled mixture and a composition with \afe=+0.4  is -0.2 mag at 10\,Gyr, in agreement with the behaviour of $\alpha$-enhanced single star population synthesis by \citet{Vazdekis15}. 

For continuously star forming populations, we examined the ultraviolet line indices studied by \citet{Calabro21} and find that some of these indices are sensitive to \afe\ while some are not. Measuring a combination of these indices, such as the 1501 and the 1719\,\AA\ indices could provide discriminating power to measure $Z$ and \afe\ simultaneously. The indices which are found to be $\alpha$-sensitive are not necessarily those with strong $\alpha$ element lines present, as the effect of iron depletion on the strength of the continuum proves to be a more significant factor in many cases. The lack of sensitivity to \afe\ in the ultraviolet in a continuously star forming population is due to the most significant contribution to the UV flux being from the youngest, hottest, most massive stars, whose individual spectra can be seen to be considerably more sensitive to \feh\ than \afe.


\section*{Acknowledgements}

We thank members of the \bpass\ team for helpful discussions. We are grateful to Charlie Conroy and Ben Johnson for providing us with a copy of the C3K spectral library. ERS and CMB acknowledge funding from the UK Science and Technology Facilities Council (STFC) through Consolidated Grant ST/T000406/1. JJE acknowledges funding from the Royal Society Te Apar\=angi of New Zealand Marsden Grant Scheme. This work made use of the University of Warwick Scientific Computing Research Technology Platform (SCRTP) and Astropy\footnote{\url{https://www.astropy.org/}}, a community-developed core Python package for Astronomy \citep{astropy:2013,astropy:2018}. 

\section*{Data Availability}

\bpass\ outputs for each metallicity and $\alpha$-enhancement value is downloadable from the \bpass\ website (\url{www.warwick.ac.uk/bpass} or \url{www.bpass.auckland.ac.nz}). At present we are making available calculations with a single IMF.
This limited release is BPASS v2.3.


\vspace*{-12pt}
\bibliographystyle{mnras}
\bibliography{mybib} 


\bsp	
\label{lastpage}







\end{document}